\documentclass{JHEP3}

\title{Giants and loops in $\beta$-deformed theories}

\author{Emiliano Imeroni and Asad Naqvi\\
Department of Physics, Swansea University\\
Singleton Park, Swansea SA2 8PP, United Kingdom\\
E-mail: \email{e.imeroni@swansea.ac.uk}, \email{a.naqvi@swansea.ac.uk}}

\abstract{We study extended objects in the gravity dual of the \Ne{1} $\beta$-deformation of \Ne{4} Super Yang--Mills theory. We identify probe brane configurations corresponding to giant gravitons and Wilson loops. In particular we identify a new class of objects, given by D5-branes wrapped on a two-torus  with a world-volume gauge field strength turned on along the torus. These appear when the deformation parameter assumes a rational value and the gauge theory spectrum has additional branches of vacua. We give an interpretation of the new D5-brane dual giant gravitons in terms of rotating vacuum expectation values in these additional branches.}

\preprint{SWAT/06/482}

\usepackage{amsmath}
\usepackage{graphicx}

\DeclareMathOperator{\tr}{Tr} 
\DeclareMathOperator{\diag}{diag} 
\newcommand{\Ne}[1]{\ensuremath{\mathcal{N}={#1}}} 
\newcommand{\gs}{\ensuremath{{g_s}}} 
\newcommand{\ads}{{\ensuremath{AdS_5\times S^5}}} 
\newcommand{\M}{\mathcal{M}}
\newcommand{\gym}{{\ensuremath{g_{\text{YM}}^2}}} 

\newcommand{\ttheta}{{\tilde{\theta}}}
\newcommand{\tphi}{{\tilde{\phi}}}
\newcommand{\tvarphi}{{\tilde{\varphi}}}
\newcommand{\txi}{{\tilde{\xi}}}
\newcommand{\tX}{{\tilde{X}}}
\newcommand{\tG}{{\tilde{G}}}
\newcommand{\tB}{{\tilde{B}}}
\newcommand{\tC}{{\tilde{C}}}
\newcommand{\hgamma}{{\hat{\gamma}}}

\begin{document}

\section{Introduction}\label{s:intro}

In the presence of Ramond-Ramond flux, a point-like graviton with a large angular momentum can sometimes expand into a spherical D-brane. This realizes the intuition that increasing the energy of the state in a theory of gravity should cause the state to expand, leading to a mixing between the UV and the IR degrees of freedom. Such
extended objects have been important in understanding the AdS/CFT correspondence~\cite{Maldacena:1998re,Gubser:1998bc,Witten:1998qj}. While single trace operators with small quantum numbers in \Ne{4} $U(N)$ Super Yang--Mills (SYM) theory are mapped onto string or particle states in the dual string theory on \ads, it has been shown in many contexts that D-branes are the natural objects to consider when quantum numbers become  of order $N$. 

In particular, gravitons propagating in \ads{} can expand into ``giant gravitons''. Giant gravitons are BPS D3-branes wrapping (in the half BPS case) a three-sphere inside the $S^5$~\cite{McGreevy:2000cw} or the $AdS_5$~\cite{Grisaru:2000zn,Hashimoto:2000zp} part of the geometry. The latter are called ``dual giant gravitons''. All of these configurations are topologically trivial, but they are stabilized by dynamics, since the D3-branes are spinning on the five-sphere.

The study of giant gravitons has led to remarkable results on the gauge/gravity correspondence. In one of the recent developments, BPS states of certain charges and preserving $1/8$ of the original supersymmetries have been counted from both sides of the correspondence by making use of giant gravitons~\cite{Kinney:2005ej,Biswas:2006tj,Mandal:2006tk}. Extending these calculations to 1/16 BPS states is of particular interest because of the presence of supersymmetric black holes with finite horizon area \cite{Gutowski:2004yv}. In another development, in \cite{Lin:2004nb}  all non-singular $1/2$ BPS solutions of type IIB supergravity preserving $SO(4) \times SO(4) \times \mathbb{R}$ isometries were classified and interpreted as 
backreacted solutions of particular distributions of giant gravitons. This work also led to an important insight about how bulk geometry might emerge from gauge theory \cite{Berenstein:2005aa}.

 D-branes have also been recently used to study Wilson loops of \Ne{4} SYM in different representations of the gauge group. While a Wilson loop in the fundamental representation is dual to a fundamental string world-sheet ending on the boundary and extending towards the bulk of $AdS$~\cite{Maldacena:1998im,Rey:1998ik}, it has been shown that Wilson loops in the symmetric and antisymmetric representations  can be more appropriately studied via configurations of D3 and/or D5 branes with fundamental string charge dissolved on their world-volume~\cite{Drukker:2005kx,Yamaguchi:2006tq,Gomis:2006sb, Hartnoll:2006hr, Hartnoll:2006is}.

The goal of this paper is to study giant gravitons and Wilson loops in a less supersymmetric setting.  This may shed an interesting light on the physics of supersymmetric black holes in $AdS_5$. Also, it would be interesting to extend the work of \cite{Lin:2004nb} constructing geometries which can be interpreted as backreacted solutions of these less supersymmetric probes, extending the ideas of emergent geometry to a less supersymmetric cases as discussed, for instance in \cite{Berenstein:2005ek}.

We will study giant gravitons and Wilson loops in  an exactly marginal deformation of \Ne{4} SYM yielding \Ne{1} conformally invariant gauge theories~\cite{Leigh:1995ep}. These provide some of the few examples of interacting four dimensional conformal field theories that have a lagrangian description. 
In particular, the ``$\beta$-deformed'' theory is obtained by replacing the superpotential of \Ne{4} SYM, which written in terms of \Ne{1} chiral superfields reads:
\begin{equation}
	W = \tr\ (\Phi_1 \Phi_2 \Phi_3 - \Phi_1 \Phi_3 \Phi_2)\,,
\end{equation}
by:
\begin{equation}\label{bdef}
	W_{\beta} = \tr\ (e^{i\pi\beta} \Phi_1 \Phi_2 \Phi_3 - e^{-i\pi\beta} \Phi_1 \Phi_3 \Phi_2)\,,
\end{equation}
where $\beta = \gamma + \tau_s \sigma$ is a complex parameter living on a torus with complex structure $\tau_s$ related to the couplings of the gauge theory. In this paper we will always consider real values of $\beta$, namely $\beta = \gamma$.

The $\beta$-deformation preserves \Ne{1} superconformal symmetry and, besides the $U(1)_R$ R-symmetry, there is a $U(1) \times U(1)$ global symmetry given by:
\begin{equation}
\begin{split}
	U(1)_1:\qquad &(\Phi_1,\Phi_2,\Phi_3) \to (\Phi_1,e^{i\varphi_1} \Phi_2, e^{-i\varphi_1} \Phi_3)\,,\\
	U(1)_2:\qquad &(\Phi_1,\Phi_2,\Phi_3) \to (e^{-i\varphi_2} \Phi_1,e^{i\varphi_2} \Phi_2, \Phi_3)\,.
\end{split}
\end{equation}
Finally, there is also a discrete $\mathbb{Z}_3$ symmetry under cyclic permutation of the superfields.

The gravity dual of the $\beta$-deformed gauge theory was found by Lunin and Maldacena (LM)~\cite{Lunin:2005jy}, and involves the product of five-dimensional anti-de Sitter space and a deformed five-sphere. It has been shown~\cite{Frolov:2005dj} that, for real deformation parameter, this gravity solution can be obtained from \ads{} via a sequence of T-dualities and shifts of coordinates (T-duality, shift, T-duality, or simply TsT) performed on a two-torus $(\varphi_1,\varphi_2)$ inside the five-sphere. In~\cite{Frolov:2005dj} it was also shown that a sequence of such transformations can be used to obtain a three-parameter family of deformed backgrounds which do not preserve any supersymmetry and are supposedly dual to conformal non-supersymmetric marginal deformations of \Ne{4} SYM.

An analysis of giant gravitons in the three-parameter non-supersymmetric background of~\cite{Frolov:2005dj} was attempted in~\cite{deMelloKoch:2005jg}, while a study of giant gravitons in the Penrose limit of the LM solution was presented in~\cite{Hamilton:2006ri}. More recently, while we were working on this project, Pirrone~\cite{Pirrone:2006iq} considered giants and dual giants in the backgrounds of~\cite{Frolov:2005dj} and~\cite{Lunin:2005jy}, finding D3-brane giant gravitons and D3-brane dual giant gravitons which do not depend on the deformation parameters, and studying the stability of the configurations. The results of~\cite{Pirrone:2006iq} have overlap with the results of our section~\ref{s:gg}. However, as we explain below we find additional dual giant graviton configurations.

We aim to study D-brane configurations in the LM background corresponding to giant gravitons and Wilson loops systematically. Our strategy is to start from the D-brane boundary conditions describing such objects in the original undeformed \ads{} space, and to follow them carefully along the TsT transformation to obtain the appropriate boundary conditions for extended objects in the LM background. The main result of the study of the transformation is that D$p$-branes in \ads{} in some cases become D$(p+2)$-branes in the LM background, which wrap the two-torus $(\varphi_1,\varphi_2)$ the TsT acts on. These expanded branes have a world-volume gauge field strength turned on along the torus which is proportional to the inverse of the deformation parameter $\gamma$, $F = 1/\gamma$, and the quantization condition for the $U(1)$ flux then requires $\gamma$ to be rational for these objects to exist.

Alongside the D3-brane giant gravitons and dual giant gravitons, for rational $\gamma$, the LM background thus also admits D5-brane dual giant gravitons. This has a nice interpretation in terms of the gauge theory dual, since for these values of the parameters one has additional branches of vacua of the theory~\cite{Berenstein:2000hy,Berenstein:2000ux,Dorey:2004xm,Benini:2004nn}. We will find the gauge theory interpretation of our new D5-brane dual giants in terms of rotating vacuum expectation values in these branches.

In the same way, besides D3-brane and D5-brane Wilson loops that are  analogous to the ones present in the undeformed \ads{} background, we find new D5-brane Wilson loops, wrapped along a two-torus where the world-volume flux is turned on, which come from the TsT of the D3-brane Wilson loop in the original theory.

The plan of the paper is as follows. We start in section~\ref{s:LM} by reviewing the Lunin-Maldacena solution dual to the $\beta$-deformed gauge theory. Then, in section~\ref{s:TsT} we present our analysis of how the boundary conditions for open strings in \ads{} transform under TsT into boundary conditions in the LM background. Section~\ref{s:gg} is dedicated to the study of D3-brane giant gravitons and D5-brane and D3-brane dual giant gravitons, and to their gauge theory interpretation, while section~\ref{s:wl} is devoted to the study of D3 and D5-brane Wilson loops. Finally, Appendix~\ref{s:Frolov} complements the text by presenting some details about the non-supersymmetric three-parameter case.

\section{The Lunin-Maldacena solution}\label{s:LM}

The type IIB supergravity solution that is dual to the $\beta$-deformation of \Ne{4} SYM was found in~\cite{Lunin:2005jy}. In the string frame it reads:
\begin{subequations}\label{LM}
\begin{align}
	ds^2 &= ds^2_{AdS_5} + R^2 \left[  \sum_i \left(d \mu_i^2+ G \mu_i^2 d\phi_i^2\right)
		+ \hat{\gamma}^2 G \mu_1^2 \mu_2^2 \mu_3^2 \big( \sum_i d\phi_i \big)^2 
		\right]\,,\\
	e^{2\phi} &= G\,,\\
	B &= R^2 \hgamma\ G\ (\mu_1^2 \mu_2^2 d\phi_1 \wedge d\phi_2
		+ \mu_2^2 \mu_3^2 d\phi_2 \wedge d\phi_3 + \mu_3^2 \mu_1^2 d\phi_3 \wedge d\phi_1)\,,\\
	C_2 &= - 4 R^2 \hgamma\ \omega_1 \wedge (d\phi_1 + d\phi_2 + d\phi_3)\,,\\
	C_4 &= \omega_4 + 4 R^4 G\ \omega_1\wedge d\phi_1 \wedge d\phi_2 \wedge d\phi_3\,,
\end{align}
\end{subequations}
where (in units where $\alpha'=1$) $R^4 = 4 \pi \gs N$, $ds^2_{AdS_5}$ is the metric on the standard $AdS_5$ space, and:
\begin{equation}
	G^{-1} = 1 + \hat{\gamma}^2 (\mu_1^2 \mu_2^2 + \mu_2^2 \mu_3^2 + \mu_3^2 \mu_1^2)\,.
\end{equation}
$\mu_i$ and $\phi_i$ parameterize a (deformed) five-sphere so that we can write:
\begin{equation}\label{forms}
\begin{gathered}
	\mu_1 = \cos \alpha\,,\qquad
	\mu_2 = \sin \alpha \cos \theta\,,\qquad
	\mu_3 = \sin \alpha \sin \theta\,,\qquad
	\sum_i \mu_i^2 = 1\,,\\
	d \omega_1 = \cos \alpha \sin^3 \alpha \sin \theta \cos \theta d \alpha \wedge d \theta\,,\qquad
	d \omega_4 = \omega_{AdS_5}\,.
\end{gathered}
\end{equation}
The parameter $\hgamma$ appearing in~\eqref{LM} is related to the deformation parameter $\gamma$ of the gauge theory by:
\begin{equation}
	\hgamma = R^2\ \gamma\,.
\end{equation}
The metric $\tG_{\mu\nu}$ and RR four-form $\tC_4$ of the undeformed \ads{} solution can obviously be found from~\eqref{LM} by putting $\gamma=0$, which implies $G=1$ (from here on, we will always use tildes to refer to coordinates and quantities in the undeformed background).

\section{TsT of D-branes}\label{s:TsT}

Our goal in the following sections is to study D-brane probes  in the LM background. A technique that we find particularly useful is to consider a specific D-brane in the undeformed background, and to map it to a configuration in the deformed one. This can be made explicit because the deformed solution~\eqref{LM} can be obtained via a ``TsT'' transformation of the original \ads{} solution~\cite{Frolov:2005dj}.

We now briefly describe how this procedure works. Start with the part of the five-sphere parameterized by $\tphi_1$, $\tphi_2$, $\tphi_3$ in the \ads{} solution and redefine coordinates in the following way:
\begin{equation}\label{varphi}
	\tphi_1 = \tvarphi_3 - \tvarphi_2\,,\qquad
	\tphi_2 = \tvarphi_3 + \tvarphi_1 + \tvarphi_2\,,\qquad
	\tphi_3 = \tvarphi_3 - \tvarphi_1\,.
\end{equation}
Acting with a T-duality along $\varphi_1$ takes to a type IIA solution, whose relevant coordinates we call $\tilde{\tvarphi}_i$. In this background, perform a shift $\tilde{\tvarphi}_2 \to \tilde{\tvarphi}_2 + \hgamma \tilde{\tvarphi}_1$. Finally, come back to a type IIB background by performing another T-duality along $\tilde{\tvarphi}_1$. Using the rules of T-duality for background fields~\cite{Bergshoeff:1995as},  we see that the resulting background is exactly given by~\eqref{LM} (for details of this procedure we refer the reader to~\cite{Frolov:2005dj}).

Since our interest resides in studying D-branes in the LM background, we need to understand how boundary conditions for open strings behave under the TsT transformation. Recall that a T-duality transformation along a direction $X^1$ of a background with metric $G_{\mu\nu}$ and NSNS two-form $B_{\mu\nu}$ yields the following boundary condition for the T-dual coordinate $\tX^1=1/X^1$:
\begin{equation}\label{Tdualbc}
	\epsilon^{\alpha\beta} \partial_\beta \tX^1
		= \eta^{\alpha\beta} G_{1a} \partial_\beta X^a
		- \epsilon^{\alpha\beta} B_{1a} \partial_\beta X^a
\end{equation}
where $\alpha,\beta=(\tau,\sigma)$ are coordinates on the world-sheet, $\eta_{\alpha\beta}$ is the world-sheet metric and we define the antisymmetric symbol by $\epsilon^{\tau\sigma}=+1$. Using~\eqref{Tdualbc} twice and taking into account the effect of the shift, we then find a relation between the coordinates of the undeformed background and the ones of the deformed background. Such a relation is most easily expressed in terms of the original coordinates $\phi_i$ of~\eqref{LM}, and can be conveniently written as:
\begin{equation}\label{TsTbc}
	\tG_{\tphi_i \tphi_j} \partial_\alpha \tphi_j
		= G_{\phi_i \phi_j} \partial_\alpha \phi_j
		- \eta_{\alpha\beta} \epsilon^{\beta\kappa} B_{\phi_i \phi_j} \partial_\kappa \phi_j\,,
\end{equation}
where $i,j=1,2,3$ and we remind that quantities with a tilde refer to the undeformed \ads{} solution.

We are now ready to study D-branes. We will concentrate on the only relevant directions, namely $\phi_i$ (or $\varphi_i$), since the remaining coordinates on the $S^5$, as well as all the ones in the $AdS_5$ part of the spacetime, are not touched by the TsT transformation. In the original (undeformed) background (where $\tB_{\mu\nu}$ vanishes and we consider D-branes without any world-volume flux turned on along the $\tphi_i$ directions), we can simply write the boundary conditions for an open string along a direction $\tphi_i$ as follows:
\begin{equation}\label{bcundef}
\begin{split}
	\partial_\sigma \tphi_i &= 0\qquad \text{(Neumann)}\\
	\partial_\tau \tphi_i &= 0\qquad \text{(Dirichlet)} \,.
\end{split}
\end{equation}
In the deformed background~\eqref{LM}, the NSNS two-form is non-vanishing and we will also allow for a non-vanishing world-volume flux $F_{ab}$.%
\footnote{Here and in the following, we will reabsorb a factor of $2\pi$ inside the world-volume field-strength $F$.}
Therefore the conditions read:
\begin{equation}\label{bcdef}
\begin{split}
	G_{\phi_i \phi_j} \partial_\sigma \phi_j
		+ ( B_{\phi_i \phi_j} - F_{\phi_i \phi_j} )\ \partial_\tau \phi_j &= 0
		\qquad \text{(mixed Neumann)}\\
	\partial_\tau \phi_i &= 0\qquad \text{(Dirichlet)}\,.
\end{split}
\end{equation}

Our aim is to map D-branes in the undeformed background, characterized by the boundary conditions~\eqref{bcundef} to D-branes in the deformed background, characterized by the conditions~\eqref{bcdef}, by making use of the transformation rules~\eqref{TsTbc} that we have derived. Let us consider the four possible cases separately. Our findings are summarized in table~\ref{t:bc}.

\TABLE{
\begin{tabular}{|c|c|c|}
\hline
\ads{} b.c. & Lunin-Maldacena b.c. & world-volume flux\\
\hline 
NNN & NNN & $F=0$\\
NND & NND & $F=0$\\
NDD & NNN & $F=1/\gamma$\\
DDD & DNN & $F=1/\gamma$\\
\hline
\end{tabular}\caption{Summary of how open string boundary conditions along $\tphi_i$ (or $\tvarphi_i$) in \ads{} are mapped onto boundary conditions along $\phi_i$ (or $\varphi_i$) in the Lunin-Maldacena background. N denotes Neumann and D denotes Dirichlet. We also indicate the world-volume field strength $F$ turned on when D directions are mapped onto N directions.\label{t:bc}}}

\begin{enumerate}

\item\emph{NNN conditions}\\
Consider a Neumann boundary condition along $\tphi_i$ in the undeformed background. We immediately see that the transformation~\eqref{TsTbc} takes the Neumann boundary condition $\partial_\sigma \tphi_i = 0$ of~\eqref{bcundef} into:
\begin{equation}\label{Ndef}
	G_{\phi_i \phi_j} \partial_\sigma \phi_j + B_{\phi_i \phi_j} \partial_\tau \phi_j = 0\,.
\end{equation}
Comparing with~\eqref{bcdef}, we see that this is the appropriate boundary condition for a D-brane extended along $\phi_i$ with no world-volume flux turned on, $F_{\phi_i \phi_j}=0$. Therefore, if we start with Neumann boundary conditions in all of the $\tphi_i$ directions in the undeformed background, we end up with all Neumann boundary conditions along the $\phi_i$ directions in the LM background, and no world-volume flux.

\item\emph{DNN conditions}\\
A Dirichlet boundary condition along $\tphi_1$ yields after the transformation:
\begin{equation}\label{Ddef}
	B_{\phi_1 \phi_j} \partial_\sigma \phi_j + G_{\phi_1 \phi_j} \partial_\tau \phi_j = 0\,,
\end{equation}
which is not immediately recognizable as either Neumann or Dirichlet in~\eqref{bcdef}. However, we see that using the Neumann boundary conditions~\eqref{Ndef} for $\phi_2$ and $\phi_3$ into~\eqref{Ddef}, the latter reduces to a Dirichlet boundary condition $\partial_\tau \phi_1=0$. Therefore, DNN in the undeformed background is mapped to DNN in the deformed one, with no flux.

\item\emph{NDD conditions}\\
The situation becomes more interesting when we have two Dirichlet boundary conditions in the undeformed background. The resulting system of equations:
\begin{equation}
\begin{cases}
	G_{\phi_1 \phi_j} \partial_\sigma \phi_j + B_{\phi_1 \phi_j} \partial_\tau \phi_j = 0\\
	B_{\phi_2 \phi_j} \partial_\sigma \phi_j + G_{\phi_2 \phi_j} \partial_\tau \phi_j = 0\\
	B_{\phi_3 \phi_j} \partial_\sigma \phi_j + G_{\phi_3 \phi_j} \partial_\tau \phi_j = 0
\end{cases}\,,
\end{equation}
using the explicit forms of $G$ and $B$, can be reduced to:
\begin{equation}
\begin{cases}
	G_{\phi_1 \phi_j} \partial_\sigma \phi_j + B_{\phi_1 \phi_j} \partial_\tau \phi_j = 0\\
	G_{\phi_2 \phi_j} \partial_\sigma \phi_j
		+ ( B_{\phi_2 \phi_j} - \frac{R^2}{\hgamma} )\ \partial_\tau \phi_j = 0\\
	G_{\phi_3 \phi_j} \partial_\sigma \phi_j
		+ ( B_{\phi_3 \phi_j} + \frac{R^2}{\hgamma} )\ \partial_\tau \phi_j = 0
\end{cases}\,,
\end{equation}
which implies that NDD conditions in the undeformed background transform into all Neumann conditions in the deformed one. The world-volume of the resulting D-brane has increased its dimension by two and now wraps a two-torus spanned by $\phi_2$ and $\phi_3$, and in addition there is a world-volume flux along the torus directions:
\begin{equation}
	F_{\phi_2 \phi_3} = \frac{R^2}{\hgamma} = \frac{1}{\gamma}\,.
\end{equation}
Notice that this magnetic world-volume field is turned on along a compact two-torus, and must then obey a quantization condition. If we put $N$ objects with these boundary conditions, the requirement is that $N/\gamma$ is integer, which is satisfied if $\gamma$ is rational, $\gamma=m/n$. In the case where $\gamma$ is not rational, the only possibility to map the object initially satisfying NDD boundary conditions onto the deformed background is by  placing it at a special point where the would-be Neumann directions are shrinking so that we effectively ``lose'' two world-volume directions.

\item\emph{DDD conditions}\\
The case of all Dirichlet boundary conditions is more easily studied in the coordinates $\varphi_i$ introduced in~\eqref{varphi}.  The situation is analogous to the NDD case we just studied, where the two Dirichlet directions $\varphi_1$ and $\varphi_2$ become Neumann after the transformation, while $\varphi_3$ remains Dirichlet:
\begin{equation}
\begin{cases}
	G_{\varphi_1 \varphi_j}\ \partial_\sigma \varphi_j
		+ ( B_{\varphi_1 \varphi_j} - \frac{R^2}{\hgamma} )\ \partial_\tau \varphi_j = 0\\
	G_{\varphi_2 \varphi_j}\ \partial_\sigma \varphi_j
		+ ( B_{\varphi_2 \varphi_j} + \frac{R^2}{\hgamma} )\ \partial_\tau \varphi_j = 0\\
	\partial_\tau \varphi_3 = 0
\end{cases}\,.
\end{equation}
One can see that the resulting brane wraps the torus now parameterized by $(\varphi_1, \varphi_2)$, with world-volume field strength:
\begin{equation}
	F_{\varphi_1 \varphi_2} = \frac{R^2}{\hgamma} = \frac{1}{\gamma}\,.
\end{equation}
The same consideration about the quantization of the $U(1)$ world-volume field that we did in the case of NDD conditions holds here as well. The resulting configuration in the deformed background makes sense only for rational values of $\gamma$, while for more generic value the brane must sit at special points where the would-be Neumann directions shrink and the brane world-volume effectively loses two directions. This point will be important for the analysis of the following sections. 
\end{enumerate}

\section{Giant gravitons}\label{s:gg}

We are now ready to study extended objects in the LM background~\eqref{LM}. Our strategy will be to consider known objects in the undeformed \ads{} background and map them to objects in the deformed one via the rules discussed in section~\ref{s:TsT}. This will give us the embedding equations which are necessary in order to study the probe action of a D$p$-brane:%
\footnote{The signs of $B$ and of the RR fields in the action~\eqref{wv} are chosen in order to be consistent with the conventions of~\cite{Lunin:2005jy,Frolov:2005dj}.}
\begin{equation}\label{wv}
	S_{\text{D}p} = - \tau_p \int d^{p+1}\sigma \ e^{-\Phi}
		\sqrt{- \det \left(\hat{G}_{ab} + F_{ab} - \hat{B}_{ab}\right) }
		 - \tau_p \int_{\mathcal{M}_{p+1}} \sum_q \hat{C}_q \wedge e^{F-\hat{B}}\,,
\end{equation}
where $F_{ab}$ is the world-volume gauge field (including a factor of $2\pi$), $\tau_p = \frac{1}{(2 \pi)^p \gs}$, and hats denote pullbacks of bulk fields onto the world-volume $\mathcal{M}_{p+1}$ of the brane. 

In this section, we will choose coordinates in the $AdS_5$ part of the spacetime in~\eqref{LM} so that:
\begin{subequations}\label{AdSmetricgg}
\begin{align}
	ds^2_{AdS_5} &= -\left(1 + \frac{r^2}{R^2}\right) dt^2 + \frac{dr^2}{1 + \frac{r^2}{R^2}}
		+ r^2 \left( d\alpha_1^2 + \sin^2\alpha_1
		\left( d\alpha_2^2 + \sin^2\alpha_2\ d\alpha_3^2 \right) \right)\,,\\
	\omega_4 &= \frac{r^4}{R} \sin^2\alpha_1 \sin\alpha_2\
		dt \wedge d\alpha_1 \wedge d\alpha_2 \wedge d\alpha_3\,.
\end{align}
\end{subequations}

\subsection{D3-brane giants}\label{s:D3g}

A giant graviton in \ads{} is a D3-brane extended along $(\ttheta, \tphi_2, \tphi_3)$ and spinning on the five-sphere: $\tphi_1 = f(\tilde{t})$\,. The boundary conditions in the $\tphi_i$ directions can be written as: 
\begin{equation}\label{gg1}
	\partial_\tau (\tphi_1 - f) = 0\,,\qquad
	\partial_\sigma \tphi_2 = 0\,,\qquad
	\partial_\sigma \tphi_3 = 0\,,
\end{equation}
and we are therefore in the NND situation of section~\ref{s:TsT}. One can easily see that the ``boosted'' boundary conditions on $\tphi_1$ does not change the result and maps onto the analogous condition $\partial_\tau (\phi_1 - f) = 0$ in the deformed background, with the \emph{same} function $f(t)$, while the conditions on $\phi_2$ and $\phi_3$ reduce to the appropriate (mixed) Neumann conditions with no world-volume field strength turned on. We therefore use the following embedding:
\begin{equation}\label{embedD3g}
\begin{gathered}
	t = \sigma^0\,,\qquad
	\theta = \sigma^1\,,\qquad
	\phi_2 = \sigma^2\,,\qquad
	\phi_3 = \sigma^3\,,\\
	\phi_1 = f (t)\,,\qquad
	\alpha = \text{const}\,,\qquad
	r = 0
\end{gathered}
\end{equation}
($r$ being the radial direction inside $AdS_5$, see~\eqref{AdSmetricgg}). The relevant part of the action~\eqref{wv} is:
\begin{equation}\label{wvD3}
	S_{\text{D}3} = - \tau_3 \int d^{4}\sigma e^{-\Phi} \sqrt{- \det ( \hat{G}_{ab} - \hat{B}_{ab} )}
		- \tau_3 \int_{\mathcal{M}_4} ( \hat{C}_4 - \hat{C}_2 \wedge \hat{B} )\,.
\end{equation}
All pieces of the computation of the DBI part combine nicely together giving:
\begin{equation}
\begin{split}
	\sqrt{- \det ( \hat{G}_{ab} - \hat{B}_{ab} )} 
		= G^{1/2} R^3 \sin^3 \alpha \sin\theta \cos\theta \sqrt{1 - R^2 \cos^2\alpha \dot{f}^2 }\,.
\end{split}
\end{equation}
Notice that the $G^{1/2}$ factor coming from the determinant precisely cancels the $G^{-1/2}$ factor coming from the dilaton, so that the DBI part of the probe reproduces \emph{exactly} the one in the undeformed case. Passing to the WZ part, we first compute:
\begin{equation}\label{wzc4c2b2}
\begin{split}
	C_4 - C_2 \wedge B 
	&= 4 R^4 \left( \omega_4 
	+ G \left(1 + \hat{\gamma}^2 (\mu_1^2 \mu_2^2 + \mu_2^2 \mu_3^2 + \mu_3^2 \mu_1^2)\right)
	\omega_1 \wedge d\phi_1 \wedge d\phi_2 \wedge d\phi_3 \right) \\
	&= 4 R^4 \left( \omega_4 
	+ \omega_1 \wedge d\phi_1 \wedge d\phi_2 \wedge d\phi_3 \right) \,.
\end{split}
\end{equation}
We then see that the resulting expression formally reduces to the four-form $\tC_4$ in the undeformed \ads{} case. Notice that this is true even before the pullback is taken, which implies that any D3-brane probe computation (with zero world-volume field strength) in the LM background will have a WZ part which gives exactly the same result as a probe with the same embedding in the undeformed case. We then see that the whole probe computation gives the same result as in the undeformed case, as was also found in~\cite{Pirrone:2006iq}. After integrating over the angular variables, we obtain the lagrangian:
\begin{equation}\label{D3gglagr}
	\mathcal{L} = \frac{N}{R} \left[ -\sin^3\alpha \sqrt{ 1 - R^2 \cos^2 \alpha \dot{f}^2} 
		+ R \sin^4 \alpha \dot{f} \right]\,.
\end{equation}
This is the same expression as the one found in~\cite{McGreevy:2000cw,Grisaru:2000zn} for the \ads{} case, and is of course solved by the same function $f(t)$ characterizing the giant graviton in the undeformed background. We are now going to review this briefly. The conserved momentum conjugate to $\phi_1$ is:
\begin{equation}
	J_{\phi_1} = N \left[ \frac{R \sin^3\alpha \cos^2\alpha \dot{f}}{\sqrt{1-R^2\cos^2\alpha \dot{f}^2}}
		+ \sin^4\alpha\right]\,,
\end{equation}
and inverting this relation we get:
\begin{equation}
	\dot{f} = \frac{J_{\phi_1} - N \sin^4\alpha}
		{R \cos\alpha \sqrt{J_{\phi_1}^2 - 2 J_{\phi_1} N \sin^4\alpha + N^2 \sin^6\alpha}}\,.
\end{equation}
The corresponding Hamiltonian is:
\begin{equation}
	\mathcal{H} = \frac{1}{R \cos\alpha} \sqrt{J_{\phi_1}^2 - 2 J_{\phi_1}
		N \sin^4\alpha + N^2 \sin^6\alpha} ~.
\end{equation}
We find degenerate minima at $\sin \alpha = 0$ and $\sin^2 \alpha = \frac{J_{\phi_1}}{N}$, both with energy $E = \frac{J_{\phi_1}}{R}$ and $f(t) = {\frac{t}{R}}$. The giant graviton corresponds to  the  $\sin^2 \alpha = \frac{J_{\phi_1}}{N}$ solution. We then obtain the bound $J_{\phi_1}  \leq N$, a manifestation of the ``stringy exclusion principle'' in the $\beta$-deformed theory. The relation between energy and momentum for these solutions indicates that these configurations preserve half of the supersymmetries of the $\beta$-deformed background. 

It is worth noting that the computation we just presented goes through unchanged even in the non-supersymmetric three-parameter solution introduced in~\cite{Frolov:2005dj}. The giant graviton configuration seems to be completely independent of supersymmetry (and of the deformation parameters). More details about the non-supersymmetric case are given in appendix~\ref{s:Frolov}.

\subsection{D5-brane dual giants}\label{s:D5dg}

We now pass to the study of dual giant gravitons, namely D3-branes extending into the $AdS$ part of the \ads{}, instead of the sphere, and still spinning on the $S^5$. One might naively expect that such a configuration, which does not extend along the directions where TsT acts, is left untouched by the transformation, but as we saw in section~\ref{s:TsT} this need not be the case. Let us start with a generic configuration in the undeformed space, where the brane is spinning along all of the $\tphi_i$ directions, so that the following DDD boundary conditions are imposed:
\begin{equation}\label{gg2}
	\partial_\tau (\tphi_1 - f_1(t)) = 0\,,\qquad
	\partial_\tau (\tphi_2 - f_2(t))= 0\,,\qquad
	\partial_\tau (\tphi_3 - f_3(t))= 0\,.
\end{equation}
These configurations of dual giants have been explicitly considered in~\cite{Mandal:2006tk}. We expect that the boundary conditions~\eqref{gg2} transform into NND conditions, yielding a D5-brane with a world-volume magnetic flux. However, we have to be careful because in this case acting with TsT on the above conditions may result in the velocity being reinterpreted as an electric component of the world-volume field. In fact, a careful analysis analogous to the one of section~\ref{s:TsT} shows that after the transformation the resulting D5-brane is extended along the directions $\varphi_1$ and $\varphi_2$ (it is natural here to switch to the coordinates defined in~\eqref{varphi}), while spinning along $\varphi_3$. Notice that the D5-brane spins along $\varphi_3 = \frac{1}{3} (\phi_1+\phi_2+\phi_3)$ independently of the direction along which the original D3-brane in the undeformed background was spinning. We can summarize our findings for the boundary conditions and electric and magnetic world-volume field components as follows:
\begin{equation}
\begin{gathered}
	\partial_\tau \left(\varphi_3 - \tfrac{f_1(t)+f_2(t)+f_3(t)}{3} \right) = 0\,,\qquad
	F_{\varphi_1 \varphi_2} = \frac{R^2}{\hgamma}\,,\\
	F_{t \varphi_1} = -\frac{R^2}{3\hgamma} (2\dot{f}_1(t)-\dot{f}_2(t)-\dot{f}_3(t)) \,,\qquad
	F_{t \varphi_2} = -\frac{R^2}{3\hgamma} (\dot{f}_1(t)+\dot{f}_2(t)-2\dot{f}_3(t))\,.
\end{gathered}
\end{equation}
We therefore choose the embedding as:
\begin{equation}\label{embedD5g}
\begin{gathered}
	t = \sigma^0\,,\quad
	\alpha_1 = \sigma^1\,,\quad
	\alpha_2 = \sigma^2\,,\quad
	\alpha_3 = \sigma^3\,,\quad
	\varphi_1 = \sigma^4\,,\quad
	\varphi_2 = \sigma^5\,,\quad
	r = \text{const}\\
	\varphi_3 = \tfrac{f_1+f_2+f_3}{3}\,,\quad
	F_{45} = \tfrac{R^2}{\hgamma}\,,\quad
	F_{04} = -\tfrac{R^2 (2\dot{f}_1-\dot{f}_2-\dot{f}_3)}{3\hgamma} \,,\quad
	F_{05} = -\tfrac{R^2 (\dot{f}_1+\dot{f}_2-2\dot{f}_3)}{3\hgamma}\,.
\end{gathered}
\end{equation}
The relevant part of the action~\eqref{wv} now reads:
\begin{multline}\label{D5wv}
	S_{\text{D}5} = - \tau_5 \int d^{6}\sigma \ e^{-\Phi}
		\sqrt{- \det \left(\hat{G}_{ab} + F_{ab} - \hat{B}_{ab}\right) }\\
		 - \tau_5 \int_{\mathcal{M}_{6}}
		 ( \hat{C}_6 + \hat{C}_4 \wedge (F - \hat{B})
		 + \hat{C}_2 \wedge (F - \hat{B}) \wedge (F - \hat{B}))\,.
\end{multline}
The computation of the determinant in the DBI part gives:
\begin{multline}\label{D5dgDBI}
	\sqrt{- \det \left(\hat{G}_{ab} + F_{ab} - \hat{B}_{ab}\right) }\\
		=  G^{1/2} \frac{R^2}{\hgamma} r^3 \sin^2\alpha_1 \sin \alpha_2
		\sqrt{1+\frac{r^2}{R^2}
		-R^2 \left(\mu_1^2 \dot{f}_1^2 + \mu_2^2 \dot{f}_2^2 + \mu_3^2 \dot{f}_3^2\right)}\,,
\end{multline}
and we see that also in this case, as in the case of the D3-brane giant of the previous section, the $G^{1/2}$ factor cancels against the contribution of the dilaton. Passing to the WZ part, it was shown in~\cite{Lunin:2005jy} that in the background~\eqref{LM} one has $C_6 - C_4 \wedge B = 0$, so that the computation reduces to:
\begin{equation}
	\tau_5 \int_{\mathcal{M}_6} (\hat{C}_4 - \hat{C}_2 \wedge \hat{B}) \wedge F
		= \tau_5 \int dt d\alpha_1 d\alpha_2 d\alpha_3 d\varphi_1 d\varphi_2
		\frac{R}{\hgamma} r^4 \sin^2\alpha_1 \sin\alpha_2\,.
\end{equation}
Putting all the pieces together (and remembering that the appropriate sign to choose for the WZ part is now the one describing an \emph{antibrane} rather than a brane~\cite{Grisaru:2000zn}), after integration over the angular variables we get the following lagrangian:
\begin{equation}\label{D5gglagrangles}
	\mathcal{L} = \frac{1}{\gamma} \frac{N}{R^4} \left[ -r^3
		\sqrt{ \Delta } 
		+ \frac{r^4}{R} \right]\,,\qquad
		\Delta = 1+\frac{r^2}{R^2}
		-R^2 \left(\mu_1^2 \dot{f}_1^2 + \mu_2^2 \dot{f}_2^2 + \mu_3^2 \dot{f}_3^2\right)\,.
\end{equation}
We see that the lagrangian has an overall dependence on the deformation parameter $\gamma$. What is the meaning of this result? As we briefly discussed in section~\ref{s:TsT}, the presence of a world-volume magnetic flux $F = 1/\gamma$ along the two-torus $(\varphi_1,\varphi_2)$ implies that this configuration is consistent only when the quantization condition for the $U(1)$ world-volume field on the D5-brane is satisfied. Generically, if we consider $N_5$ of these D5-branes, the quantization condition requires $N_5/\gamma$ to be an integer, which is the case if $\gamma = m/n$, with $m$ and $n$ coprime, and $N_5$ a multiple of $m$. These rational values of $\gamma$ are precisely the ones where new chiral primary states and branches of vacua of the gauge theory appear, so we expect to be able to interpret our dual giant graviton states in terms of these, see section~\ref{s:ggfield}.

In any case, we now proceed with the study of the conserved energy and momenta of the solutions to the equations of motion following from~\eqref{D5gglagrangles}. It is convenient to rewrite $\Delta$ in~\eqref{D5gglagrangles} in terms of $\varphi_3$ and of the world-volume field strength by inverting the relations in~\eqref{embedD5g}:
\begin{equation}\label{delta2}
	\Delta = 1+\frac{r^2}{R^2}
		-R^2 \left(\mu_1^2 (\dot{\varphi}_3-\gamma F_{04})^2
		+ \mu_2^2 (\dot{\varphi}_3+\gamma F_{04}-\gamma F_{05})^2
		+ \mu_3^2 (\dot{\varphi}_3+\gamma F_{05})^2\right)\,.
\end{equation}
Now recall that $\varphi_1=\sigma^4$ and $\varphi_2=\sigma^5$ are world-volume coordinates of the D5-brane, so that we can compute the conserved currents in the original DBI lagrangian using $j_a = \frac{\partial\mathcal{L}_{\text{DBI}}}{\partial F_{0b}} F_{ba}$. In terms of the one-dimensional lagrangian~\eqref{D5gglagrangles}-\eqref{delta2} we then obtain the conserved charges:
\begin{equation}\label{dgmomentavarphi}
	J_{\varphi_1} = \frac{\partial\mathcal{L}}{\partial F_{05}} F_{54}\,,\qquad
	J_{\varphi_2} = \frac{\partial\mathcal{L}}{\partial F_{04}} F_{45}\qquad\text{and}\quad
	J_{\varphi_3} = \frac{\partial\mathcal{L}}{\partial \dot{\varphi}_3}\,.
\end{equation}
The resulting expressions are easily expressed in terms of the original $\phi_i$ coordinates. Using:
\begin{equation}
	J_{\phi_1} = \tfrac{1}{3} J_{\varphi_1} - \tfrac{2}{3} J_{\varphi_2} + \tfrac{1}{3} J_{\varphi_3}\,,\quad
	J_{\phi_2} = \tfrac{1}{3} J_{\varphi_1} + \tfrac{1}{3} J_{\varphi_2} + \tfrac{1}{3} J_{\varphi_3}\,,\quad
	J_{\phi_3} = - \tfrac{2}{3} J_{\varphi_1} + \tfrac{1}{3} J_{\varphi_2} + \tfrac{1}{3} J_{\varphi_3}\,,
\end{equation}
we get:
\begin{equation}\label{dgmomenta}
	J_{\phi_i} = \frac{N r^3 \mu_i^2 \dot{f}_i}{\gamma R^2 \sqrt{\Delta}}
		= \frac{\partial\mathcal{L}}{\partial \dot{f}_i} \qquad (i=1,2,3).
\end{equation}
The conserved momenta depend on the $\mu_i$, namely on the point in $(\alpha,\theta)$ where the dual giant is sitting.

Proceeding with the computation of the Hamiltonian, we find (see also~\cite{Kim:2005mw} for an analogous computation of giant graviton probes with electric fields in \ads):
\begin{equation}\label{dgH}
\begin{split}
	\mathcal{H} &= J_{\varphi_3} \dot{\varphi}_3 + \frac{\partial\mathcal{L}}{\partial{F_{0a}}} F_{0a} - \mathcal{L}
		= \frac{\partial\mathcal{L}}{\partial \dot{f}_i} \dot{f}_i - \mathcal{L}\\
		&= \frac{1}{\gamma} \frac{N}{R}
		\left[ \sqrt{\left(1+\frac{r^2}{R^2} \right)
		\left( \frac{\gamma^2}{N^2} \sum_{i=1}^3 \frac{J_{\phi_i}^2}{\mu_i^2}
		+\frac{r^6}{R^6} \right)} - \frac{r^4}{R^4} \right]\,.
\end{split}
\end{equation}

We now want to concentrate on the BPS solutions. In the \ads{} background, it was shown that the dual giant graviton~\eqref{gg2} generically preserves $1/8$ of the original supersymmetries if the condition
\begin{equation}\label{dualsusy}
	f_1(t) = f_2(t) = f_3(t) = f(t)
\end{equation}
is satisfied~\cite{Mandal:2006tk}. In the case at hand, imposing~\eqref{dualsusy} we see that the electric components of the field strength on the D5-brane world-volume vanish. The Hamiltonian~\eqref{dgH} reduces to:
\begin{equation}\label{dgH2}
	\mathcal{H} = \frac{1}{\gamma} \frac{N}{R}
		\left[ \sqrt{\left(1+\frac{r^2}{R^2} \right)
		\left( \frac{\gamma^2 J_{\varphi_3}^2}{N^2}+\frac{r^6}{R^6} \right)} - \frac{r^4}{R^4} \right]\,,
\end{equation}
with
\begin{equation}
	J_{\varphi_3} = J_{\phi_1} + J_{\phi_2} + J_{\phi_3}
		= \frac{N r^3 \dot{f}}{\gamma R^2 \sqrt{1+\frac{r^2}{R^2}-R^2 \dot{f}^2}}\,.
\end{equation}
Extremizing~\eqref{dgH2}, we find minima at $r=0$ and $\frac{r^2}{R^2} = \frac{\gamma J_{\varphi_3}}{N}$, with energy $E = \frac{J_{\varphi_3}}{R} = \frac{J_{\phi_1} + J_{\phi_2} + J_{\phi_3}}{R}$, as appropriate for a BPS state. 

We now argue that, in this case where $\gamma = \frac{m}{n}$ is rational, the momenta $J_{\varphi_1}$ and $J_{\varphi_2}$ have to be multiples of $n$ in the quantum theory.  Because of the presence of the magnetic field $F_{45}$, the coordinates $\varphi_1$ and $\varphi_2$ do not commute with each other: $[ \hat{\varphi}_1, \hat{\varphi}_2]=\frac{1}{F_{45}} = \gamma$. When $\gamma= \frac{m}{n}$, the theory on the noncommutative torus parameterized by $(\varphi_1, \varphi_2)$ is equivalent to another theory on an ordinary commutative torus, with the identifications:
\begin{equation}\label{momquant}
	\varphi_1 \rightarrow \varphi_1 + \frac{2\pi}{n}\,,\qquad
	\varphi_2 \rightarrow \varphi_2 + \frac{2\pi}{n}\,.
\end{equation}
This can be seen as follows. Functions on a non-commutative torus can be expanded as:
\begin{equation}
	g(\hat{\varphi}) = \sum_{\vec{J} \in \mathbb{Z}^2} g_{\vec{J}}
		~ e^{i \vec{J}\cdot \vec{\hat{\varphi}}}\,,
\end{equation}
where $\vec{\hat{\varphi}}$ is the vector $(\hat{\varphi}_1, \hat{\varphi}_2)$.  It is easy to show that~\cite{Saraikin:2000dn}:
\begin{equation}
	[e^{i n \vec{J} \cdot \vec{\hat{\varphi}}}, g(\hat{\varphi})]=0\,,
\end{equation}
which implies that we can treat the exponents $e^{i \vec{J}\cdot \vec{\hat{\varphi}}}$ as ordinary functions on a commutative torus if $J_{\varphi_1}=J_{\varphi_2} = 0\! \mod n$, which in turn implies the periodicity conditions \eqref{momquant}.%
\footnote{An alternative argument using T-duality can be found in section 6 of \cite{Seiberg:1999vs}.} From this, we find the following condition that must be satisfied by the momenta $J_{\phi_i}$ of the D5-brane dual giant:
\begin{equation}\label{D5momenta}
	J_{\phi_1}=  J_{\phi_2}= J_{\phi_3} \mod n\,.
\end{equation}

\subsection{D3-brane dual giants}\label{s:D3dg}

In the previous sections, we learnt that generically a D3-brane which does not extend on the five-sphere is mapped onto a D5-brane with two world-volume directions along the deformed $S^5$ of the LM background. However, it is still a legitimate question to ask whether a D3-brane dual giant which only extends on the $AdS$ part of the spacetime exists.

We can then attempt a probe computation starting with the following embedding for a D3-brane, directly analogous to the one used in the undeformed case, see~\eqref{gg2}:
\begin{equation}\label{embedD3dg}
\begin{gathered}
	t = \sigma^0\,,\quad
	\alpha_1 = \sigma^1\,,\quad
	\alpha_2 = \sigma^2\,,\quad
	\alpha_3 = \sigma^3\,,\\
	r = \text{const}\,,\quad
	\phi_1 = f_1(t)\,,\quad
	\phi_2 = f_2(t)\,,\quad
	\phi_3 = f_3(t)\,.
\end{gathered}
\end{equation}

The computation of the determinant appearing in the DBI part of the action~\eqref{wvD3} now yields a complicated expression, which in particular does not cancel the $G^{-1/2}$ factor coming from the dilaton. Also, it does not reproduce the structure inside the square root in~\eqref{D5dgDBI}, which was  a feature both of the D3-brane dual giant probe in the undeformed background and of the D5-brane probe of the previous section.

However, we can choose values of $(\alpha,\theta)$ corresponding to the usual choices one makes in \ads{} where the brane only spins along one of the $\tphi_i$ directions:
\begin{subequations}\label{3cases}
\begin{align}
	&\text{(i)}& \mu_1&=1\,,& & \mu_2=\mu_3=0 &  &\left(\alpha=0\right)\,,\\
	&\text{(ii)}& \mu_2&=1\,,& & \mu_3=\mu_1=0& &\left(\alpha=\tfrac{\pi}{2}\,,\:\: \theta=0\right)\,,\\
	&\text{(iii)}& \mu_3&=1\,,& & \mu_1=\mu_2=0&
		&\left(\alpha=\tfrac{\pi}{2}\,,\:\: \theta=\tfrac{\pi}{2}\right)\,.
\end{align}
\end{subequations}
In such cases, one has $G=1$ and the full expression simplifies. Concentrating on the BPS case~\eqref{dualsusy} where $f_1(t) = f_2(t) = f_3(t) = f(t)$, in the end we obtain the lagrangian 
\begin{equation}\label{D3dgglagr}
	\mathcal{L} = \frac{N}{R^4} \left[ -r^3 \sqrt{ 1 + \frac{r^2}{R^2} - R^2 \dot{f}^2} 
		+ \frac{r^4}{R} \right]\,,
\end{equation}
with no dependence on the deformation parameter $\gamma$. This is unlike the result~\eqref{D5gglagrangles}  for the D5-brane dual giant. The computation of the solutions and their energies is analogous to the one in the previous sections. In terms of the conserved momentum
\begin{equation}\label{D3dgJ}
	J_{\varphi_3} = \frac{N r^3 \dot{f}}{R^2 \sqrt{1+\frac{r^2}{R^2}-R^2 \dot{f}^2}}\,,
\end{equation}
the Hamiltonian is:
\begin{equation}\label{D3dgH}
	\mathcal{H} = \frac{N}{R}
		\left[ \sqrt{\left(1+\frac{r^2}{R^2} \right)
		\left( \frac{J_{\varphi_3}^2}{N^2}+\frac{r^6}{R^6} \right)} - \frac{r^4}{R^4} \right]\,.
\end{equation}
Therefore, in each of the three cases of~\eqref{3cases} we obtain two solutions, $r=0$ and $\frac{r^2}{R^2} = \frac{J_{\varphi_3}}{N}$. The energy is  $E=\frac{J_{\varphi_3}}{R}=\frac{J_{\phi_1}+J_{\phi_2}+J_{\phi_3}}{R}$, where the momenta $(J_{\phi_1},J_{\phi_2},J_{\phi_3})$ in the three cases~\eqref{3cases} are respectively given by $(J_{\varphi_3},0,0)$, $(0,J_{\varphi_3},0)$ and $(0,0,J_{\varphi_3})$.

\subsection{Gauge theory interpretation}\label{s:ggfield}

In this section, we will make some remarks about the field theory interpretation of the giant gravitons we obtained in the previous sections. 

The gauge theory dual to the Lunin-Maldacena background \eqref{LM} is the \Ne{1} super-conformal theory with three chiral superfields $\Phi_i$ ($i=1,2,3$) and a superpotential  
\begin{equation}\label{bdef1}
	W_{\beta} = \tr\ (e^{i\pi\beta} \Phi_1 \Phi_2 \Phi_3 - e^{-i\pi\beta} \Phi_1 \Phi_3 \Phi_2)\,.
\end{equation}
This is a deformation of \Ne{4} SYM by an exactly marginal operator. 

Chiral primary operators of this theory correspond to BPS states in the gravity dual. For generic values of $\beta$, there are single trace chiral primary operators $\tr (\Phi_1^{J_1} \Phi_2^{J_2}\Phi_3^{J_3})$ with:
\begin{equation}\label{chir1}
	(J_1,J_2,J_3)=(J,0,0)\,,\quad (0,J,0)\,,\quad (0,0,J)\,,\quad (J,J,J)\,,
\end{equation}
where $J$ is an integer. When $\beta$ is real and rational, $\beta=\gamma=\frac{m}{n}$,   there are additional single trace operators with:
\begin{equation}\label{chir2}
	J_1 = J_2 = J_3 \mod n\,.
\end{equation}

Single trace chiral primaries are dual to single particle BPS states in the bulk description. When the quantum numbers $J_i$ become large, the relevant objects to consider in the gravity dual are instead D-brane giant gravitons, which map to operators that are no longer single trace~\cite{Balasubramanian:2001nh,Corley:2001zk}. In the case at hand, we expect that the giant gravitons studied in section~\ref{s:gg} will be dual to gauge theory operators whose quantum numbers match the ones presented above for single trace chiral primaries.

Both the D3-brane giant of section~\ref{s:D3g} and the D3-brane dual giant of section~\ref{s:D3dg} exist for any value of $\gamma$, and it is thus natural to expect that they are dual to operators whose quantum numbers belong to the branch~\eqref{chir1}.  We start  with the D3-brane giant graviton of section~\ref{s:D3g}. The angular momentum  of the giant spinning along $\phi_1$ was of type $(J, 0, 0)$ and we found no dependence on the deformation parameter. We propose that the dual gauge theory operator is precisely the same ``subdeterminant'' operator which is dual to the usual giant graviton in \ads~\cite{Balasubramanian:2001nh}:
\begin{equation}\label{gdualop}
	\mathcal{O}_{J} = \frac{1}{J!}
		\epsilon_{a_1 \dotsm a_J c_{J+1} \dotsm c_{N}}
		\epsilon^{b_1 \dotsm b_J c_{J+1} \dotsm c_{N}}
		(\Phi_1)^{a_1}_{\phantom{a_1}b_1} \dotsm (\Phi_1)^{a_J}_{\phantom{a_J}b_J}\,.
\end{equation}
As also discussed in~\cite{Hamilton:2006ri,Pirrone:2006iq}, the fact that the operator is unchanged with respect to the undeformed case is justified because the structure of~\eqref{gdualop} implies that, when we replace the usual product of field by the $\star$-product appropriate for the $\beta$-deformed theory, the resulting phases are vanishing. Of course, operators analogous to~\eqref{gdualop} with $\Phi_2$ or $\Phi_3$ instead of $\Phi_1$ are dual to giant gravitons spinning along $\phi_2$ or $\phi_3$.

In the same way, we can interpret the D3-brane dual giant gravitons of section~\ref{s:D3dg} having charge $(J,0,0)$ as being dual to the operator~\cite{Corley:2001zk}:
\begin{equation}\label{dgdualop}
	\tilde{\mathcal{O}}_J = \frac{1}{J!} \sum_{\sigma \in \mathcal{S}_J}
		(\Phi_1)^{a_1}_{\phantom{a_1}a_{\sigma(1)}} \dotsm
		(\Phi_1)^{a_J}_{\phantom{a_J}a_{\sigma(J)}}\,,
\end{equation}
with $\mathcal{S}_J$ the permutation group of length $J$, as in the \Ne{4} theory. Analogous operators are dual to the D3-brane dual giants with charge $(0,J,0)$ and $(0,0,J)$. Notice that the branch~\eqref{chir1} also contains $(J,J,J)$ states, but we have not found D3-brane giants or dual giants with such quantum numbers.

We now want to find the gauge theory interpretation of the new D5-brane dual giant gravitons found in section~\ref{s:D5dg}, whose quantum numbers~\eqref{D5momenta} match the operators in the branch~\eqref{chir2}. In order to discuss this case, we first have to introduce an alternative gauge theory interpretation of dual giant gravitons introduced in~\cite{Hashimoto:2000zp}. There, the dual giant graviton which expanded into a D3-brane in the $AdS_5$ part of the original $AdS_5 \times S^5$ geometry was interpreted as a classical rotating vacuum expectation value in the \Ne{4} theory. This is a  classical solution of the \Ne{4} equations of motion in which one of the three complex scalars has the following form:
\begin{equation}
	\Phi \sim \Phi_0 e^{i\omega t}\,,
\end{equation}
where $\Phi_0= \diag \left(v, -\frac{v}{N-1}, \cdots , -\frac{v}{N-1}\right)$. The $\mathcal{N}=4$ theory defined on the $S^3$ does not have a Coulomb branch because of the conformal coupling to the curvature of $S^3$. However, these classical solutions are the analogs of the Coulomb branch that exists when the theory lives on $\mathbb{R}^4$. 

We will look for similar rotating vacuum expectation values for the $\beta$-deformed theory on $S^3 \times \mathbb{R}$ and identify them with the dual giant gravitons we have found, in particular with the new dual giants of section~\ref{s:D5dg}. The part of the lagrangian that includes the three complex scalars is:
\begin{equation}
	L_{\beta} = -\frac{\sqrt{-g}}{\gym} \tr \left( \partial_\mu \bar{\Phi}_i \partial^\mu \Phi^i
		+ \frac{1}{R^2} \bar{\Phi}_i \Phi^i
		+ \frac{1}{4} \left[ \Phi_i,  \bar{\Phi}^i \right] \left[ \Phi_j, \bar{\Phi}^j \right] 
		- \frac{1}{2} \left[ \Phi_i,  \Phi_j \right]_\beta \left[ \bar{\Phi}^i, \bar{\Phi}^j \right]_\beta\right)\,,
\end{equation}
where the $\beta$-deformed commutator is defined as:
\begin{equation}
	\left[ \Phi_i, \Phi_j \right]_{\beta} = e^{i \pi\beta} \Phi_i \Phi_j - e^{-i \pi \beta} \Phi_j \Phi_i\,.
\end{equation}
Just like \Ne{4} SYM, the $\beta$-deformed theory on $S^3$ does not have a Coulomb branch because of the conformal mass term for the scalars. However, the Coulomb branch that would exist on $\mathbb{R}^4$ gets mapped to rotating vacuum expectation values, precisely as we described above for the case of the \Ne{4} theory.

The vacuum structure of the $\beta$-deformed theory differs qualitatively from that of \Ne{4} theory. For generic values of $\beta$, the would-be Coulomb branch of vacua, where the $\Phi_i$ are diagonal and only one of them is non-zero, gets mapped to rotating vevs which are analogous to the ones just described for the \Ne{4} theory. These rotating vevs correspond to the D3-brane dual giants of section~\ref{s:D3dg}, which we have also matched to the operators~\eqref{dgdualop}.

However, for rational values of $\beta = \gamma = \frac{1}{n}$, the theory has extra Higgs branches where the rank of the unbroken group is $k= \frac{N}{n}$, as  was discussed in~\cite{Dorey:2003pp}. We will now analyze the rotating vevs corresponding to these branches for the theory on $S^3$.

For the three scalar fields, we will put in the ansatz for spatially uniform, but time dependent background values of the form:
\begin{equation}\label{phivevs}
	\Phi_1 = e^{i \phi_1(t)} \M^{(1)} \otimes U_{(n)}\,,\qquad
	\Phi_2 = e^{i \phi_2(t)} \M^{(2)} \otimes V_{(n)}\,,\qquad
	\Phi_3 =  e^{i \phi_3(t)} \M^{(3)} \otimes W_{(n)}\,,
\end{equation}
where $\M^{(i)}$ are three diagonal $k \times k$ matrices:
\begin{equation}
	\M^{(i)}= \diag \left(v^i, -\frac{v^i}{k-1}, \cdots , -\frac{v^i}{k-1} \right)\,,
\end{equation}
$U_{(n)}$ and $V_{(n)}$ are the usual ``clock'' and ``shift'' $n \times n$ matrices:
\begin{equation}
 	U_{(n)} = \begin{pmatrix}
		1 & & &\\
 		& e^{\frac{2\pi i}{n}} & & \\
		& & \ddots & \\
		& & & e^{\frac{2\pi i}{n} (n-1)} 
		\end{pmatrix}\,,\qquad
	V_{(n)} = \begin{pmatrix}
		0 & 1 & & & \\
		& 0 & 1 & &\\
		\phantom{\ddots} & \phantom{\ddots} & \ddots & \ddots & \phantom{\ddots} \\
		& & & 0 & 1 \\
		1 & & & & 0  
		\end{pmatrix}\,,
\end{equation}
and $W_{(n)}= V^\dagger_{(n)} U^\dagger_{(n)}$. It is useful to define $\eta^i$ as follows:
\begin{equation}
	\eta^i= \sqrt{\frac{\gym N}{\Omega_3 R^2}} v^i\,,
\end{equation}
where $\Omega_3$ is the volume of a three-sphere of unit radius. By substituting the form~\eqref{phivevs} of $\Phi_i$ in the lagrangian, we get an effective lagrangian for the dynamical variables $\eta_i(t)$ and $\phi_i(t)$:
\begin{equation}
	L= \xi\ \frac{N R}{2} \sum_i \left(\dot{\eta}_i^2+ \eta_i^2 \dot{\phi}_i ^2 - \frac{\eta_i^2}{R^2} \right)\,.
\end{equation}
Here, the prefactor $\xi$ comes from the overall trace in front of the lagrangian and is given by
$\xi = n ( 1+ \frac{1}{k-1})$.  

Defining the three conserved momenta as $J_i = \frac{\partial L}{\partial {\dot{\phi}_i}}$ , the Hamiltonian of the system is (for constant $\eta_i(t)$):
\begin{equation}
	H=  J_i \dot{\phi}_i -L
		= \sum_{i=1}^{3} \left( 
		 \frac{1}{\xi} \frac{J_i^2}{2 N R \eta_i^2} + \xi \frac{N \eta_i^2}{2R} \right)\,.
\end{equation}
This is minimized when $\eta_i^2 = \frac{J_i}{\xi N}$. At the minimum, the energy is given by:
\begin{equation}
	E= \frac{J_1 + J_2 +J_3}{R}\,.
\end{equation}

So far, our analysis has been purely classical. In the quantum theory, the momenta $J_i$ will be integers since the angles $\phi_i$ have period $2 \pi$. Moreover, we will now argue that the $J_i$ have in fact to be equal modulo $n$. There is a redundancy in the parameterization of the rotating vevs as defined in \eqref{phivevs}, given by the gauge transformations:
\begin{equation}
	\Phi_i \rightarrow g_a \Phi_i g_a^\dagger\,,
\end{equation}
for $i=1,2,3$ and $a=1,2$, with $g_1= I_{(m)} \otimes U_{(n)}$ and $g_2= I_{(m)} \otimes V_{(n)}$. Under these transformations, the variables $\phi_i$ transform as:
\begin{equation}
\begin{aligned}
	g_1&: & \phi_1 &\to \phi_1\,, &\quad \phi_2 &\to \phi_2 - \frac{2 \pi}{n}\,,
		&\quad \phi_3 &\to \phi_3 + \frac{2\pi}{n}\,,\\
	g_2&: & \phi_1 &\to \phi_1 + \frac{2\pi}{n}\,, & \phi_2 &\to \phi_2\,,
		& \phi_3 &\to \phi_3 - \frac{2 \pi}{n}\,.
\end{aligned}
\end{equation}
These identifications are more usefully rewritten in terms of linear combinations of $\phi_i$ defined as:
\begin{equation}
	\varphi_1= \frac{ \phi_1 + \phi_2 -2 \phi_3}{3}\,,\qquad
	\varphi_2= \frac{ -2 \phi_1 + \phi_2 + \phi_3}{3}\,,\qquad
	\varphi_3 = \frac{ \phi_1 + \phi_2 + \phi_3}{3}\,.
\end{equation}
Notice that these are the field theory counterparts of the coordinates defined in~\eqref{varphi} that we have been using throughout our study of probes in the LM background. In terms of the $\varphi_i$, the gauge transformations act as:
\begin{equation}
	g_1\,: \quad \varphi_1 \to \varphi_1 - \frac{2\pi}{n}\,,\qquad\quad
	g_1 \ g_2\,: \quad \varphi_2 \to \varphi_2 - \frac{2\pi}{n}\,.
\end{equation}
This implies that in the quantum theory the momenta conjugate to $\varphi_1$ and $ \varphi_2$ should be multiples of $n$, i.e. $J_{{\varphi}_1}=J_{{\varphi}_2}= 0\! \mod n$. This in turn implies that the momenta $J_i$ for these configurations have to be equal modulo $n$:
\begin{equation}
	J_1= J_2 =J_3 \mod n\,.
\end{equation}

We then see that these classical solutions of the $\beta$-deformed gauge theory on $S^3 \times \mathbb{R}$, corresponding to rotating vacuum expectation values, have all the right features to be the dual of the D5-brane dual giant gravitons we have considered in section~\ref{s:D5dg}. In particular, they match precisely their quantum numbers (including the  quantization conditions) and energy.

The D5-brane dual giant gravitons are the compact counterparts of the D5 branes with world-volume $\mathbb{R}^{1,3} \times T^2$ found in   \cite{Dorey:2003pp,Dorey:2004iq} . In that case, the boundary theory was defined on $\mathbb{R}^{1,3}$ which is the boundary of Poincar\'e $AdS_5$. As such, the theory had a Coulomb branch of vacua, and for rational values of $\gamma$ additional branches emerged, where the vacuum expectation values were as in~\eqref{phivevs} without any time dependent factor. The D5 branes thus provided a bulk description of these branches. 

\section{Wilson loops}\label{s:wl}

Wilson loops in certain representations of the gauge group are dual to D-brane configurations in the bulk geometry. In this section, we study such objects in the Lunin-Maldacena background. 

For  \Ne{4} SYM, the analysis performed in~\cite{Maldacena:1998im,Yamaguchi:2006tq,Gomis:2006sb} shows that for type IIB string theory on \ads{}, there are three objects preserving the same supersymmetry and the same global symmetry $SU(1,1) \times SU(2) \times SO(5)$ as the straight line Wilson loop:
\begin{itemize}\setlength{\itemsep}{0pt}
\item Fundamental string with $AdS_2$ world-sheet,
\item D5-brane with $AdS_2 \times S^4$ world-volume (where $S^4 \subset S^5$),
\item D3-brane with $AdS_2 \times S^2$ world-volume (where $S^2 \subset AdS_5$).
\end{itemize}
The first of the three configurations corresponds to the original prescription by Maldacena~\cite{Maldacena:1998im}, while the study of D-brane configurations in the context of Wilson loops has been initiated in~\cite{Drukker:2005kx}. 

The analysis of~\cite{Drukker:2005kx,Yamaguchi:2006tq,Gomis:2006sb} shows that an $AdS_2 \times S^2$ D3-brane with $k$ units of world-volume flux along the $AdS_2$ directions represents a Wilson loop in the $k$-index symmetric representation, while an $AdS_2 \times S^4$ D5-brane with the same flux represents a Wilson loop in the $k$-index antisymmetric representation. The proposal of~\cite{Gomis:2006sb} also includes Wilson loops in arbitrary representations, which can be described as collections of either D3 or D5-branes. From this perspective, the expectation value of the loop is computed by evaluating the action of a probe D-brane in the \ads{} background. 

Our goal is then to perform an analogous study of D-brane configurations which are dual to Wilson loops in (anti)symmetric representations of the gauge group in the $\beta$-deformed gauge theory. The strategy we follow will be the same as in section~\ref{s:gg}, namely ``following'' the relevant D-branes of the undeformed \ads{} background through the TsT transformation which leads to the LM background, making use of the results we presented in section~\ref{s:TsT}.

We now summarize our results.  We  find that a Wilson loop in the gravity dual of the $\beta$-deformed theory with $k$ units of fundamental string charge can be described by the following D-branes:
\begin{itemize}\setlength{\itemsep}{0pt}
\item D5-brane with $AdS_2 \times \tilde{S}^4$ world-volume (where $\tilde{S}^4$ is a deformed four-sphere inside the deformed $S^5$),
\item D5-brane with $AdS_2 \times S^2 \times T^2$ world-volume (where $S^2 \subset AdS_5$ and there is a world-volume flux $F=\frac{1}{\gamma}$ along the $T^2 \subset S^5$. These configurations only exist for rational values of $\gamma$),
\item D3-brane with $AdS_2 \times S^2$ world-volume (where $S^2 \subset AdS_5$).
\end{itemize}
As in the undeformed \ads{} case, all of the above D-brane configurations have a world-volume flux along the $AdS_2$ part of the world-volume that gives the appropriate fundamental string charge.

These D-brane configurations are found to preserve two $U(1)$ symmetries, given by appropriate combinations of the $U(1)_R \times U(1)_1 \times U(1)_2$ symmetry of the theory. The main result we find is that the D-brane probes reproduce the same results one finds in \ads{} for Wilson loops in \Ne{4} SYM. However, we also find the additional D5-brane configuration wrapped on a two-torus with flux, which again, as was the case for the new dual giants of section~\ref{s:D5dg}, appear when the $\beta$-deformed gauge theory has additional branches of vacua. It would be interesting to study the gauge theory interpretation of these objects more carefully, and we plan to return to it in the future.

Our study of Wilson loop D-brane probes will closely follow the one in~\cite{Lunin:2006xr}. Throughout this section, it will be useful to adopt the following parameterization of the $AdS_5$ part of the spacetime, which makes the $AdS_2\times S^2$ structure manifest:
\begin{subequations}\label{adscoords}
\begin{align}
	ds^2_{AdS_5} &= R^2 \left( \cosh^2 \rho\ \tfrac{-dt^2+dy^2}{y^2} + d\rho^2
		+ \sinh^2 \rho \left(d\chi_1^2 + \sin^2 \chi_1 d\chi_2^2 \right) \right)\,,\\
	d\omega_4 &= \omega_{AdS_5}
		= 4R^4 \cosh^2 \rho \sinh^2 \rho\ \tfrac{\sin \chi_1}{y^2}\
		d\rho \wedge dt \wedge dy \wedge d\chi_1 \wedge d\chi_2\,.
\end{align}
\end{subequations}

\subsection{D5-brane loops without flux}\label{s:D5w}

We start from the D5-branes with $AdS_2\times S^4$ world-volume in the undeformed \ads{} background. Notice that the parameterization of the metric~\eqref{LM} (for $\hgamma=0$)  for the five-sphere,
\begin{equation}
	ds^2_{S^5} = R^2 \left[ d\alpha^2 + \cos^2\alpha\ d\tphi_1^2+ \sin^2\alpha\ d\Omega_3^2 \right]\,,
\end{equation}
is not suited for an $S^4$ embedding. It is  useful to  perform the following change of coordinates:
\begin{equation}\label{S4change}
	\sin\alpha = \sin\beta \sin\txi\,,\qquad
	\tan\tphi_1 = \tan\beta \cos\txi\,,
\end{equation}
that puts the metric in the form:
\begin{equation}
	ds^2_{S^5} = R^2 \left[ d\beta^2 + \sin^2\beta \left( d\xi^2 + \sin^2\xi\ d\Omega_3^2 \right)\right]\,.
\end{equation}
In these coordinates, we easily see that a four-sphere is parameterized by $(\xi,\theta,\tphi_2,\tphi_3)$ for constant $\beta$. The boundary conditions for a D-brane extended along this $S^4$ are Neumann along $\tphi_2$ and $\tphi_3$, so we know from the analysis of section~\ref{s:TsT} that they will be preserved in the deformed background.

Passing then to the deformed background (and to un-tilded coordinates, with the same change of coordinates as above), the embedding of the D5-brane Wilson loop can be chosen as:
\begin{equation}\label{embedD5w}
\begin{gathered}
	t = \sigma^0\,,\qquad
	y = \sigma^1\,,\qquad
	\theta = \sigma^2\,,\\
	\xi = \sigma^3\,,\qquad
	\phi_2 = \sigma^4\,,\qquad
	\phi_3 = \sigma^5\,,\qquad
	\beta = \text{const}\,.
\end{gathered}
\end{equation}
In addition, in order to induce fundamental string charge there will be a world-volume field strength proportional to the volume of the $AdS_2$ space:
\begin{equation}
	F = E R^2\ \tfrac{dt\wedge dy}{y^2}\,.
\end{equation}

The relevant part of the action is again given by~\eqref{D5wv}. The determinant in the DBI part gives:
\begin{equation}
	\sqrt{- \det \left(\hat{G}_{ab} + F_{ab} - \hat{B}_{ab}\right) }
		= G^{1/2} \frac{R^6}{y^2} \sin^4\beta \sin^3\xi \sin\theta \cos\theta \sqrt{\cosh^4\rho-E^2}\,,
\end{equation}
and we see that the $G^{-1/2}$ factor coming from the dilaton in the DBI action is again canceled, as in the case of the D3-brane giant graviton in section~\ref{s:gg}.

Moving on to the  WZ part, recall from section~\ref{s:gg} that, since $C_6 - C_4 \wedge B$ is zero in the background~\eqref{LM}, the relevant coupling is the one to $C_4 - C_2 \wedge B$. This expression, as shown in~\eqref{wzc4c2b2}, reduces formally to the four-form $\tilde{C}_4$ in the undeformed  background. Thus we only need to express $\tilde{C}_4$ in the new coordinates~\eqref{S4change}.  Since its expression is simpler, let us compute the expression of the five-form $\tilde{F}_5$ instead:
\begin{equation}
	\tilde{F}_5 = 4R^4 \sin^4\beta \sin^3\xi \sin\theta \cos\theta
		d\theta \wedge d\beta \wedge d\phi_2 \wedge d\phi_3\,.
\end{equation}
This is all we need to compute the WZ contribution:
\begin{equation}
	\tau_5 \int_{\mathcal{M}_6} (\hat{C}_4 - \hat{C_2} \wedge \hat{B}) \wedge F
		= \tau_5 \int dt dy d\theta d\beta d\phi_2 d\phi_3 \frac{4 E R^6}{y^2}
		\sin^3\xi \sin\theta \cos\theta \int^\beta dx \sin^4 x\,.
\end{equation}
The complete probe action then becomes:
\begin{equation}
	S_{\text{D5}} = \tau_5 R^6 \int d^6 \sigma
		\sin^3 \xi \sin\theta \cos\theta
		\left[- \sin^4\beta \sqrt{\cosh^4\rho - E^2} + 4 E \int^\beta dx \sin^4 x\right]\,,
\end{equation}
and we see that we have  obtained exactly the same result as in the undeformed \ads{} case (compare for instance the expression in~\cite{Lunin:2006xr}). Fixing $\rho=0$ to preserve the $S^2$ symmetry, minimization with respect to $\beta$ yields the equation:
\begin{equation}
	-4 \sin^3\beta \left(\sqrt{1-E^2} \cos\beta - E \sin\beta \right) = 0\,,
\end{equation}
which has two solutions, $\beta=0$ and $\cos \beta = E$.

\subsection{D5-brane loops with flux}\label{s:D35w}

We now proceed to the fate of the $AdS_2\times S^2$ D3-brane Wilson loops in the undeformed background. Such branes are characterized by all Dirichlet boundary conditions for the open string along $\tvarphi_i$ (we will work in the coordinates $\varphi_i$ of~\eqref{varphi}, and unlike in the previous section we will revert to the parameterization~\eqref{LM} of the (deformed) five-sphere part of the geometry). After the TsT transformation, the configuration is mapped to a D5-brane wrapped on the $(\varphi_1,\varphi_2)$ torus with world-volume field strength $F_{\varphi_1\varphi_2} = R^2/\hgamma = 1/\gamma$.

We  then choose the embedding:
\begin{equation}\label{embedD35w}
\begin{gathered}
	t = \sigma^0\,,\qquad
	y = \sigma^1\,,\qquad
	\chi_1 = \sigma^2\,,\qquad
	\chi_2 = \sigma^3\,,\\
	\varphi_1 = \sigma^4\,,\qquad
	\varphi_2 = \sigma^5\,,\qquad
	\rho = \text{const}\,,
\end{gathered}
\end{equation}
with world-volume field strength:
\begin{equation}
	F_{ty} = \frac{ER^2}{y^2}\,,\qquad
	F_{\varphi_1\varphi_2} = \frac{R^2}{\hgamma}\,.
\end{equation}
With this embedding, the determinant in the DBI part of the action~\eqref{wv} for a D5-brane gives:
\begin{equation}
	\sqrt{- \det \left(\hat{G}_{ab} + F_{ab} - \hat{B}_{ab}\right) }
		= G^{1/2} \frac{R^6}{\hgamma y^2} \sinh^2\rho \sin\chi_1 \sqrt{\cosh^4\rho-E^2}\,,
\end{equation}
so that  the $G^{1/2}$ cancels with the $G^{-1/2}$ factor from the dilaton. Using the form of $F_5$ deduced from~\eqref{adscoords}, the WZ contribution gives:
\begin{equation}
	\tau_5 \int_{\mathcal{M}_6} (\hat{C}_4 - \hat{C_2} \wedge \hat{B}) \wedge F
		= \tau_5 \int dt dy d\chi_1 d\chi_2 d\varphi_1 d\varphi_2\
		\frac{4 R^6}{\hgamma y^2} \sin\chi_1 \int^\rho dx \cosh^2 x \sinh^2 x\,.
\end{equation}
Combining the DBI and WZ contribution, and performing the integration along the angular coordinates $\varphi_1$ and $\varphi_2$, we get the final result for the probe action:
\begin{equation}\label{D35w}
	S_{\text{D5}} = \tau_3 \frac{R^4}{\gamma} \int dt dy d\chi_1 d\chi_2
		\frac{\sin \chi_1}{y^2}
		\left[- \sinh^2\rho \sqrt{\cosh^4\rho - E^2} + 4  \int^\rho dx \cosh^2 x \sinh^2 x \right]\,.
\end{equation}
The resulting action has the same functional form as the one of the D3-brane Wilson loop in the undeformed background (compare for instance with~\cite{Lunin:2006xr}) but, as was the case for the D5-brane dual giant in section~\ref{s:gg}, differs from it by an overall factor $1/\gamma$. Because of quantization of magnetic flux on the torus, these configurations are only allowed for rational values of $\gamma$. 
As in the \ads{} case, the action~\eqref{D35w} has two solutions, $\rho=0$ and $\cosh\rho = E$.

\subsection{D3-brane loops}\label{s:D3w}

As we did when studying dual giant gravitons in section~\ref{s:gg}, we can look for a D3-brane configuration which more closely resembles the D3-brane Wilson loop of the undeformed theory. Choosing the same embedding one would choose in the \ads{} case:
\begin{equation}\label{embedD3w}
\begin{gathered}
	t = \sigma^0\,,\qquad
	y = \sigma^1\,,\qquad
	\chi_1 = \sigma^2\,,\qquad
	\chi_2 = \sigma^3\,,\qquad
	\rho = \text{const}\,,
\end{gathered}
\end{equation}
one sees that the computation of the D3-brane probe action~\eqref{wvD3} takes us to a situation which parallels the case of D3-brane dual giant of section~\ref{s:D3dg}: the determinant in the DBI part does not cancel the $G^{-1/2}$ factor coming from the dilaton, so that the extremization of the action with respect to $\alpha$ and $\theta$ requires the D3-brane probe to sit at one of the special points where two of the three circles parameterized by the $\phi_i$ shrink so that $G=1$. These are the cases, listed in~\eqref{3cases} of section~\ref{s:D3dg}, where two of the $\mu_i$ vanish. The action then reduces to the same one finds in the undeformed case:
\begin{equation}\label{D3w}
	S_{\text{D3}} = \tau_3 R^4 \int dt dy d\chi_1 d\chi_2
		\frac{\sin \chi_1}{y^2}
		\left[- \sinh^2\rho \sqrt{\cosh^4\rho - E^2} + 4  \int^\rho dx \cosh^2 x \sinh^2 x \right]\,.
\end{equation}
which is also the same action~\eqref{D35w} that we found for D5-brane Wilson loops with flux, except for the overall dependence on the deformation parameter $\gamma$.

\acknowledgments

We would like to thank Sujay Ashok, Paolo Di Vecchia, Tim Hollowood, Jihye Seo, and especially Prem Kumar and Carlos N\'u\~nez for valuable discussions. We also thank the fourth Simons Workshop held at Stony Brook in August 2006 for providing a stimulating environment where part of this work was done. The work of AN is supported by a PPARC advanced fellowship.

\appendix

\section{Frolov's non-supersymmetric solution}\label{s:Frolov}

The probe computations we report in the main text hold more or less unchanged in the three-parameter solution found by Frolov~\cite{Frolov:2005dj}. This non-supersymmetric solution is obtained by performing three TsT transformations starting from the \ads{} solution, acting on the $(\phi_1,\phi_2)$ torus with parameter $\hgamma_3$, then on the $(\phi_2,\phi_3)$ torus with parameter $\hgamma_1$ and finally on the $(\phi_3,\phi_1)$ torus with parameter $\hgamma_2$. The solution reads:
\begin{subequations}\label{Frolov}
\begin{align}
	ds^2 &= ds^2_{AdS_5} + R^2 \left[  \sum_i \left(d \mu_i^2+ G \mu_i^2 d\phi_i^2\right)
		+ G \mu_1^2 \mu_2^2 \mu_3^2 \big( \sum_i \hgamma_i d\phi_i \big)^2 
		\right]\,,\\
	e^{2\phi} &= G\,,\\
	B &= R^2 G\ (\hgamma_3 \mu_1^2 \mu_2^2 d\phi_1 \wedge d\phi_2
		+ \hgamma_1 \mu_2^2 \mu_3^2 d\phi_2 \wedge d\phi_3
		+ \hgamma_2 \mu_3^2 \mu_1^2 d\phi_3 \wedge d\phi_1)\,,\\
	C_2 &= - 4 R^2 \omega_1 \wedge (\hgamma_1 d\phi_1 +\hgamma_2  d\phi_2 +\hgamma_3  d\phi_3)\,,\\
	C_4 &= \omega_4 + 4 R^4 G\ \omega_1\wedge d\phi_1 \wedge d\phi_2 \wedge d\phi_3\,,
\end{align}
\end{subequations}
where we use the expressions in~\eqref{forms} and where:
\begin{equation}
	G^{-1} = 1 + (\hgamma_3^2 \mu_1^2 \mu_2^2 +\hgamma_1^2  \mu_2^2 \mu_3^2 +\hgamma_2^2  \mu_3^2 \mu_1^2)\,.
\end{equation}
The supersymmetric solution~\eqref{LM} of~\cite{Lunin:2005jy} is recovered from~\eqref{Frolov} by setting all parameters equal, $\hgamma_1=\hgamma_2=\hgamma_3=\hgamma$.

In order to study extended objects in the background~\eqref{Frolov}, one can repeat the analysis of boundary conditions we did in section~\ref{s:TsT} for the supersymmetric case, and the results are qualitatively the same, namely Neumann boundary conditions on the $\tphi_i$ in \ads{} are mapped by the TsT transformations onto Neumann boundary conditions on the $\phi_i$ in the new background, while in the case where one has two Dirichlet boundary conditions they transform into Neumann with a magnetic field on the world-volume of the brane.

One can then see that the probe computation for the D3-brane giant graviton of section~\ref{s:D3g} and the one for the D5-brane Wilson loop of section~\ref{s:D5w} (as well as the one for the dual giant and the D3-brane Wilson loop at special points presented in sections~\ref{s:D3dg} and~\ref{s:D3w}) hold unchanged with respect to the non-supersymmetric case, with no dependence on the deformation parameters.

The cases where the world-volume of the brane after TsT has acquired two additional directions behave in the same way as in the case of the supersymmetric LM solution too, but let us present few more details. We begin with the configuration analogous to the one considered in section~\ref{s:D5dg}. If we start from a D3-brane dual giant graviton in \ads{} that spins along the $\tphi_i$ directions as in~\eqref{gg2}, we can summarize the boundary conditions and world-volume gauge field strength components resulting from the TsT transformations as:
\begin{equation}\label{Frolovdgbc}
\begin{gathered}
	\partial_\tau \left(\varphi_3 - \tfrac{\hgamma_1 f_1(t)+\hgamma_2 f_2(t)+\hgamma_3 f_3(t)}{3}\right) = 0\,,\qquad
	F_{\varphi_1 \varphi_2} = \tfrac{R^2 (\hgamma_1+\hgamma_2+\hgamma_3)^2}{9\hgamma_1\hgamma_2\hgamma_3}\,,\\
	F_{t \varphi_1} = -\tfrac{R^2 (\hgamma_1+\hgamma_2+\hgamma_3)}{9\hgamma_1\hgamma_2\hgamma_3}
		(2\hgamma_1 \dot{f}_1(t)-\hgamma_2 \dot{f}_2(t)-\hgamma_3 \dot{f}_3(t)) \,,\\
	F_{t \varphi_2} = -\tfrac{R^2 (\hgamma_1+\hgamma_2+\hgamma_3)}{9\hgamma_1\hgamma_2\hgamma_3}
		(\hgamma_1 \dot{f}_1(t)+\hgamma_2 \dot{f}_2(t)-2\hgamma_3 \dot{f}_3(t))\,,
\end{gathered}
\end{equation}
where now we have chosen $\varphi_i$ coordinates which are related to $\phi_i$ by:%
\footnote{There is of course arbitrariness in the choice of the transformations~\eqref{Frolovvarphi}. The only relevant inputs coming from the study of the TsT transformations are that the $\phi_i$ coordinates must appear in the combination $\hgamma_i \phi_i$ (for instance because the Dirichlet direction after TsT turns out to be $\hgamma_1 \phi_1 + \hgamma_2 \phi_2 + \hgamma_3 \phi_3$), and that we want the transformation to reduce to~\eqref{varphi} in the supersymmetric case. Our choice is made in order to yield a result which is symmetric in $\hgamma_i$. Other choices yield physically equivalent results.}
\begin{equation}\label{Frolovvarphi}
	\hgamma_1 \phi_1 = \tfrac{\hgamma_1+\hgamma_2+\hgamma_3}{3} (\varphi_3-\varphi_2)\,,\quad
	\hgamma_2 \phi_2 = \tfrac{\hgamma_1+\hgamma_2+\hgamma_3}{3} (\varphi_3+\varphi_1+\varphi_2)\,,\quad
	\hgamma_3 \phi_3 = \tfrac{\hgamma_1+\hgamma_2+\hgamma_3}{3} (\varphi_3-\varphi_1)\,.
\end{equation}
The computation then proceeds  as in section~\ref{s:D5dg}, and the final result for the lagrangian is:
\begin{equation}\label{Frolovdglagr}
	\mathcal{L} = \frac{(\gamma_1+\gamma_2+\gamma_3)^2}{9\gamma_1\gamma_2\gamma_3}
		\frac{N}{R^4} \left[ -r^3 \sqrt{1+\frac{r^2}{R^2}
		-R^2 \left(\mu_1^2 \dot{f}_1^2 + \mu_2^2 \dot{f}_2^2 + \mu_3^2 \dot{f}_3^2\right)} 
		+ \frac{r^4}{R} \right]\,.
\end{equation}
Starting from this expression, the computation of the energy and momenta proceeds in analogy with the supersymmetric case. In particular, we note that the magnetic flux in~\eqref{Frolovdgbc}, combined with the periodicities one can infer from~\eqref{Frolovvarphi}, imply a quantization for the three parameters $\gamma_i$ analogous to the one of $\gamma$ in the case of the D5-branes in the LM background: Again new states seem to appear when the deformation parameters assume rational values.

Finally, one can also perform the probe computation for the D5-brane Wilson loop with world-volume flux, analogously to the one presented in section~\ref{s:D35w} for the LM background. The only difference with respect to the latter is again the use of the coordinates~\eqref{Frolovvarphi} and the expression of the magnetic flux, which is $F_{\varphi_1 \varphi_2} = \tfrac{R^2 (\hgamma_1+\hgamma_2+\hgamma_3)^2}{9\hgamma_1\hgamma_2\hgamma_3}$ as in~\eqref{Frolovdgbc}. The final result for the lagrangian is then:
\begin{multline}\label{FrolovD35w}
	S_{\text{D5}} = \tau_3 R^4 \frac{(\gamma_1+\gamma_2+\gamma_3)^2}{9\gamma_1\gamma_2\gamma_3}\\
		\times \int dt dy d\chi_1 d\chi_2
		\frac{\sin \chi_1}{y^2}
		\left[- \sinh^2\rho \sqrt{\cosh^4\rho - E^2} + 4 E \int^\rho dx \cosh^2 x \sinh^2 x \right]\,.
\end{multline}
and we again recall the need of the quantization of the deformation parameters $\gamma_i$ for this configuration to make sense.

\bibliographystyle{../bibtex/myutcaps}
\bibliography{../bibtex/mybib}

\end{document}